\documentclass[prc,preprint,superscriptaddress,showpacs,nofootinbib,tightenlines,amsmath]{revtex4}
\usepackage{graphics}
\usepackage{epsfig}
\usepackage{slashed}
\usepackage{pbox, calc}
\usepackage{slashed}

\usepackage{xparse}
\usepackage{tikz}
\usepackage{pgfpages}
\usetikzlibrary{calc,backgrounds,trees,positioning,arrows,chains,shapes.geometric,%
    decorations.pathreplacing,decorations.pathmorphing,decorations.markings,shapes,%
    matrix,shapes.symbols}
\tikzstyle{every picture}+=[remember picture] \tikzstyle{na} =
[baseline=-.5ex] \tikzstyle{background grid}=[draw,
black!50,step=.5cm]
\usetikzlibrary{calc,shapes,arrows,automata,backgrounds,calendar}
\usetikzlibrary{intersections}
\usetikzlibrary{matrix}
\usetikzlibrary{trees}
\usetikzlibrary{decorations.pathmorphing}
\usetikzlibrary{decorations.pathreplacing}
\usetikzlibrary{decorations.markings}
\usetikzlibrary{3d}
\usetikzlibrary{external}
\pgfdeclaredecoration{complete sines}{initial} {
  \state{initial}[
    width=+0pt,
    next state=sine,
    persistent precomputation={\pgfmathsetmacro\matchinglength{
        \pgfdecoratedinputsegmentlength / int(\pgfdecoratedinputsegmentlength/\pgfdecorationsegmentlength)}
        \setlength{\pgfdecorationsegmentlength}{\matchinglength pt}
    }] {}
  \state{sine}[width=\pgfdecorationsegmentlength]{
    \pgfpathsine{\pgfpoint{0.25\pgfdecorationsegmentlength}{0.5\pgfdecorationsegmentamplitude}}
    \pgfpathcosine{\pgfpoint{0.25\pgfdecorationsegmentlength}{-0.5\pgfdecorationsegmentamplitude}}
    \pgfpathsine{\pgfpoint{0.25\pgfdecorationsegmentlength}{-0.5\pgfdecorationsegmentamplitude}}
    \pgfpathcosine{\pgfpoint{0.25\pgfdecorationsegmentlength}{0.5\pgfdecorationsegmentamplitude}}
  }
  \state{final}{}
}

\tikzset{
  photon/.style=  {
                    decorate,
                    solid,
                    decoration={complete sines, amplitude=1mm, segment length=1.75mm, post length=0}
                  },
  rho/.style= {
                solid
              },
  rhoarrow/.style=  {
                  solid,
                  postaction={decorate},
                  decoration={markings, mark=at position .6 with {\arrow{>}}}
                }
}
 \tikzset{->-/.style={decoration={
  markings,
  mark=at position #1 with {\arrow{>}}},postaction={decorate}}}
\tikzset{-<-/.style={decoration={
  markings,
  mark=at position #1 with {\arrow{<}}},postaction={decorate}}}
\begin{document}
\title{Mass and width of the $\Delta(1232)$ resonance using complex-mass
renormalization}
\author{T. Bauer}
\affiliation{PRISMA Cluster of Excellence, Institut f\"ur
Kernphysik, Johannes Gutenberg-Universit\"at Mainz, D-55099 Mainz, Germany}
\author{Y. \"Unal}
\affiliation{PRISMA Cluster of Excellence, Institut f\"ur
Kernphysik, Johannes Gutenberg-Universit\"at Mainz, D-55099 Mainz, Germany}
\affiliation{Physics Department, \c{C}anakkale  Onsekiz Mart University, 17100 \c{C}anakkale, Turkey}
\author{A. K\"u\c{c}\"ukarslan}
\affiliation{Physics Department, \c{C}anakkale  Onsekiz Mart University, 17100 \c{C}anakkale, Turkey}
\author{S.~Scherer}
\affiliation{PRISMA Cluster of Excellence, Institut f\"ur
Kernphysik, Johannes Gutenberg-Universit\"at Mainz, D-55099 Mainz,
Germany}
\date{22 December 2016}
\preprint{ MITP/16-143}
\begin{abstract}
   We discuss the pole mass and the width of the $\Delta(1232)$ resonance to third order in chiral effective field theory.
   In our calculation we choose the complex-mass renormalization scheme (CMS) and show that the CMS provides
a consistent power-counting scheme.
   In terms of the pion-mass dependence, we compare the convergence behavior of the CMS
with the small-scale expansion (SSE).
\end{abstract}
\pacs{12.39.Fe,14.20.Gk}
\maketitle

\section{Introduction}

   Effective field theories (EFTs) have been successfully applied to various
sectors of the strong interactions, which are based on quantum chromodynamics
(QCD) as the underlying fundamental theory with quarks and gluons as dynamical
degrees of freedom.
   At low energies, chiral perturbation theory (ChPT)~\cite{Weinberg:1978kz,Gasser:1983yg,Gasser:1984gg}
is the EFT of QCD, making use of the Goldstone bosons of spontaneous chiral symmetry breaking
as effective dynamical degrees of freedom (see, e.g., Refs.~\cite{Ecker:1994gg,Bernard:1995dp,Scherer:2002tk,Scherer:2012zzd} for an introduction).
   The application of dimensional regularization in combination with a modified minimal subtraction scheme
results in an unambiguous correspondence between the loop expansion and the chiral expansion
\cite{Gasser:1983yg,Gasser:1984gg}, thus providing a straightforward power counting.
   In general, the inclusion of heavy and, especially, resonant degrees of freedom in
chiral EFTs leads to a more complex situation, because one has to solve
the question of implementing a systematic power-counting scheme.
   To be specific, in relativistic baryon chiral perturbation theory (RBChPT),
applying the same renormalization condition as in mesonic ChPT did not work out,
because loops containing internal nucleon lines violated the power counting
\cite{Gasser:1987rb}.
   Besides the (non-relativistic) heavy-baryon formulation of
chiral perturbation theory (HBChPT)~\cite{Jenkins:1990jv,Bernard:1992qa},
several renormalization prescriptions have been developed,
leading to a ``standard'' power counting in the covariant
framework~\cite{Tang:1996ca,Ellis:1997kc,Becher:1999he,Gegelia:1999gf,Gegelia:1999qt,Fuchs:2003qc,Schindler:2003xv}.

   The $\Delta(1232)^{++}$ resonance was first observed in $\pi^+ p$ scattering \cite{Anderson:1952nw},
and the ratio of $\pi^+$ to $\pi^-$ scattering was interpreted by K.~A.~Brueckner in terms of
an excited state with spin 3/2 and isospin 3/2 \cite{Brueckner:1952zz}.
   Ever since its discovery, the $\Delta(1232)$ resonance has played a prominent role in the
description of low and medium-energy processes such as pion-nucleon
scattering, electromagnetic pion production, Compton scattering, etc.~(see, e.g.,
Refs.~\cite{Ellis:1997kc,Hemmert:1996xg,Hemmert:1996rw,Fettes:2000bb,Pascalutsa:2002pi,Pascalutsa:2005vq,Pascalutsa:2006up,McGovern:2012ew,Blin:2016itn,Yao:2016vbz}).
   This is due to the comparatively small mass gap between the $\Delta(1232)$ and
the nucleon, the strong coupling of the $\Delta(1232)$ to the $\pi N$ channel,
and its relatively large photon decay amplitudes.
   Therefore, it would be preferable to extend BChPT by including the
$\Delta(1232)$ as an explicit degree of freedom in both the $\pi N$ threshold region and
the resonance region \cite{Pascalutsa:2002pi,Jenkins:1991es,Banerjee:1994bk,Hemmert:1997ye,Bernard:2003xf,Hacker:2005fh,Pascalutsa:2005nd,Epelbaum:2015vea}.

   In the following, we will apply the complex-mass scheme (CMS)
to the unstable $\Delta(1232)$ degrees of freedom.
   The CMS~\cite{Stuart:1990vk,Denner:1999gp,Denner:2006ic,Actis:2006rc,Actis:2008uh}
was originally designed to derive properties of $W$, $Z_0$, and Higgs bosons
obtained from resonant processes.
    The idea is to assign complex renormalized masses to the
resonances, which are defined as the complex pole positions of
the corresponding full propagators.
   In the CMS, complex gauge-boson masses are used in tree-level and loop
calculations, which then imply complex counter terms in the Lagrangian.
   The application of the CMS to hadronic resonances was for the first time considered
in Ref.~\cite{Djukanovic:2009zn}.
   In the meantime, the CMS has been used in the calculation of various
hadronic properties of unstable particles and reactions involving unstable
particles as intermediate states \cite{Yao:2016vbz,Epelbaum:2015vea,
Djukanovic:2009zn,Djukanovic:2009gt,Bauer:2012at,Djukanovic:2013mka,Bauer:2014cqa,Djukanovic:2014rua,Gegelia:2016xcw,Gegelia:2016pjm}.
   The applicability of the CMS in hadronic EFT at two-loop order was shown in Ref.~\cite{Djukanovic:2015gna}, and two-loop
calculations for the widths of the Roper resonance and the $\Delta(1232)$ resonance were
discussed in Refs.~\cite{Gegelia:2016xcw} and \cite{Gegelia:2016pjm},
respectively.
   An importrant question is whether the S matrix is
perturbatively unitary in the CMS.
   In Ref.~\cite{Bauer:2012gn}, this was explicitly shown to be the case at the one-loop level
for a model of light fermions interacting with a heavy vector boson.
   A general proof can be found in Ref.~\cite{Denner:2014zga}.
   Since perturbative unitarity of the S matrix is ensured, the CMS can be considered
as a rigorous formalism for defining a renormalized quantum field theory
\cite{Denner:2006ic,Beneke:2015vfa}.

   In the present article we discuss the mass and the width of the $\Delta(1232)$ in relativistic chiral EFT at $\mathcal{O}(q^3)$
in the framework of the CMS.
   In Sec.~II, we discuss the effective Lagrangian and the power-counting scheme.
   In Sec.~III, we calculate the complex pole position of the $\Delta(1232)$ and discuss the result
as a function of the pion mass.
   Section IV contains a short summary.

\section{Effective Lagrangian}
   The most general effective Lagrangian for the calculation of the pole position of the Delta propagator can
be written as
\begin{equation}
\mathcal{L}_{\rm eff}={\cal L}_\pi+{\cal L}_{\pi N}+{\cal L}_{\pi \Delta}+{\cal L}_{\pi N\Delta},
\label{genlag}
\end{equation}
see, e.g., Refs.~\cite{Scherer:2012zzd,Hacker:2005fh} for details on the definition of the building blocks.
   We expand the Lagrangians in terms of the pion fields and display only the expressions relevant for
the calculation of the Delta self energy at order three,
\begin{align}
{\cal L}_\pi^{(2)}&
=\frac{1}{2}\partial_\mu\vec\pi\cdot\partial^\mu\vec\pi-\frac{1}{2}M^2\vec{\pi}^2+\cdots,
\label{Lpi2}\\
{\cal L}^{(1)}_{\pi N}&
=\bar{\Psi}\left(i\slashed{\partial}-m_{N0}\right)\Psi
-\frac{1}{2}\frac{\texttt{g}_A}{F}\bar{\Psi}\gamma^\rho\gamma_5 \vec\tau\cdot\partial_\rho\vec\pi\Psi +\cdots,
\label{LpiN1}\\
{\cal L}_{\pi \Delta}^{(1)}&=-\bar{\Psi}_{\mu} \xi^{\frac{3}{2}}
\left[(i\slashed{\partial}-m_{\Delta 0})g^{\mu \nu}-i(\gamma^{\mu}\partial^{\nu}+\gamma^{\nu}\partial^{\mu})
+i\gamma^{\mu}\slashed{\partial}\gamma^{\nu}+m_{\Delta 0}\gamma^{\mu}\gamma^{\nu}\right]\Psi_{\nu}\nonumber\\
&\quad+\frac{1}{2}\frac{\texttt{g}_{A\Delta}}{F}\bar{\Psi}_{\mu} \xi^{\frac{3}{2}}
\left[\gamma^\rho\gamma_5\vec\tau\cdot\partial_\rho\vec\pi g^{\mu\nu}
-\left(\gamma^\mu\vec\tau\cdot\partial^\nu\vec\pi+\vec\tau\cdot\partial^\mu\vec\pi\gamma^\nu\right)\gamma_5
-\gamma^\mu\gamma^\rho\gamma_5\vec\tau\cdot\partial_\rho\vec\pi\gamma^\nu\right] \xi^{\frac{3}{2}}\Psi_\nu\nonumber\\
&\quad+\cdots,
\label{LpiDelta1}\\
{\cal L}_{\pi N\Delta}^{(1)}&=-\frac{1}{2}\frac{\texttt{g}_{AN\Delta}}{F}\bar{\Psi}_{\mu, i}\;\xi_{ij}^\frac{3}{2}\left(g^{\mu \nu}-\gamma^{\mu}\gamma^{\nu}\right)
\partial_\nu\pi_j\Psi+\text{h.c.}+\cdots,
\label{LpiNDelta1}\\
\mathcal{L}_{\pi\Delta}^{(2)}&=-4c_1^{\Delta}M^2\bar{\Psi}_\mu\xi^\frac{3}{2}g^{\mu \nu}\Psi_{\nu}+\cdots.
\label{LDelta2}
\end{align}
   In Eqs.~(\ref{Lpi2})--(\ref{LDelta2}), $\pi_i$ denotes the cartesian component of the pion triplet,
$\Psi$ the nucleon isospin doublet, $\Psi_{\nu,j}$ are the vector-spinor isovector-isospinor Rarita-Schwinger
fields \cite{Rarita:1941mf} of the Delta resonance, and $\xi^{\frac{3}{2}}$ is the isospin-$3/2$ projection operator.
   The superscripts $(i)$ denote the order $i$ in the chiral expansion; $M$ is the pion mass at leading order in the quark-mass expansion:
$M^2=2 B\hat m$, where $B$ is related to the quark condensate $\langle\bar{q}q\rangle_0$ in the chiral limit;
$m_{N0}$ and $m_{\Delta0}$ are the bare masses of the nucleon and the Delta; $g_A$ and $F$ refer to the chiral
limit of the axial-vector coupling constant and the pion-decay constant, respectively.
   The constants $\texttt{g}_{A\Delta}$ and $\texttt{g}_{AN\Delta}$ are the $\Delta$ axial-vector coupling constant
and the $N\Delta$-transition axial-vector coupling constant in the chiral limit, respectively (in Ref.~\cite{Scherer:2012zzd},
they are denoted by $g_1=\texttt{g}_{A\Delta}$ and $2g=\texttt{g}_{AN\Delta}$).
   Finally, $c_1^{\Delta}$ is responsible for a quark mass correction to the (complex) Delta mass and has a similar structure
as the corresponding term in the nucleon Lagrangian \cite{Gasser:1987rb,Bernard:1992qa}.

   In a Lorentz-invariant Lagrangian formulation of a field theory involving particles of higher spin ($J\geq 1$), one
necessarily introduces unphysical degrees of freedom \cite{Rarita:1941mf,Moldauer:1956zz}.
   Therefore, constraints need to be imposed, which specify the physical degrees of freedom.
   Here, we make use of the constraint analysis of Ref.~\cite{Wies:2006rv} using the canonical
(Hamilton) formalism \cite{Dirac,Gitman:1990qh,teitelboim,Djukanovic:2010tb,Unal:2015hea}, which results in a self-consistent theory
with the correct number of degrees of freedom and leads to consistent interactions of the Delta resonance with
pions and nucleons.
    As a consequence of the constraint analysis, the Lagrangian is invariant under the so-called point transformation
\cite{Wies:2006rv,Nath:1971wp,Tang:1996sq}, and we may set the so-called off-shell parameter $A$ to $-1$,
which also results in a simplified form of the Delta propagator.
   Note that our approach is somewhat different from Refs.~\cite{Yao:2016vbz,Gegelia:2016xcw,Gegelia:2016pjm},
where the Lagrangian has been obtained by applying field redefinitions to remove redundant structures
depending on off-shell parameters \cite{Scherer:1994wi,Tang:1996sq,Krebs:2008zb,Krebs:2009bf}.
   Both methods generate the same number of independent low-energy coupling constants.

   When expressing the bare Lagrangian in terms of a basic Lagrangian and a counter-term Lagrangian,
we split the bare parameters of the Lagrangians Eqs.~(\ref{Lpi2})--(\ref{LDelta2}) into renormalized parameters and counter-term
contributions:
\begin{equation}
m_{\Delta 0}=z_{\Delta}+\delta z_{\Delta},\quad
m_0=m+\delta m,\quad
\ldots,
\label{d}
\end{equation}
where $z_{\Delta}$ is the complex pole of the Delta propagator in the chiral limit,
$m$ is the mass of the nucleon in the chiral limit, and the ellipses stand for other parameters of the Lagrangian.
   We use the renormalized mass parameters in the free propagators and include the counter terms perturbatively.
   To summarize, the basic Lagrangians corresponding to Eqs.~(\ref{Lpi2})--(\ref{LDelta2})
contain in total 10 real parameters, namely, $M^2$, $F$, $m$, $z_\Delta$, $\texttt{g}_A$, $\texttt{g}_{A\Delta}$, $\texttt{g}_{AN\Delta}$,
and $c_1^\Delta$, where $z_\Delta$ and $c_1^\Delta$ are complex numbers.

   In the CMS, we implement the following power counting. A renormalized diagram is said to be of
${\cal O}(q^D)$, where $q$ denotes a small momentum or a pion mass.
   An interaction vertex derived from an ${\cal O}(q^i)$ Lagrangian [superscript $(i)$] counts as
$q^i$, a pion propagator as order $q^{-2}$, a nucleon and a Delta propagator as $q^{-1}$, and a loop
integration in $n$ dimensions as $q^n$.
   The order $D$ is then obtained from \cite{Ecker:1994gg}
\begin{align}
\label{eq:DN}
D&=n N_L-2I_\pi-I_B+\sum_{i=1}^\infty i N_i,
\end{align}
where $N_L$, $I_\pi$, $I_B$, and $N_i$ denote the number of independent loop momenta, internal
pion lines, internal baryon lines, and vertices derived from ${\cal O}(q^i)$ Lagrangians,
respectively.

\section{Pole mass and width of the $\Delta(1232)$ resonance}

     Making use of isospin symmetry, the isospin structure of the full propagator of the Delta resonance is given by
\cite{Hacker:2005fh}
\begin{equation}
S_{ij,\alpha \beta}^{\mu \nu}(p)=\xi_{ij,\alpha \beta}^\frac{3}{2} S^{\mu \nu}(p),
\label{vel}
\end{equation}
where $S^{\mu \nu}(p)$ is the solution of the equation
\begin{equation}
S^{\mu \nu}(p)=S_0^{\mu \nu}(p)-S^{\mu \rho}(p)\Sigma_{\rho \sigma}(p)S_0^{\sigma \nu}(p).
\label{c}
\end{equation}
   Here, $S_0^{\mu \nu}(p)$ stands for the free propagator and $i\Sigma^{\mu \nu}$ for the sum of all one-particle irreducible diagrams
contributing to the two-point Green function.
   The self energy of the Delta resonance is parametrized as follows \cite{Hacker:2005fh},
\begin{eqnarray}
\Sigma^{\mu \nu}(p)=\sum_{a=1}^{10}\Sigma_a(p^2)\mathcal{P}_a^{\mu \nu},
\label{cd}
\end{eqnarray}
with
\begin{equation}
\begin{aligned}
&\mathcal{P}_1^{\mu \nu}=g^{\mu \nu},\quad \mathcal{P}_2^{\mu \nu}=\gamma^{\mu} \gamma^{\nu},\quad \mathcal{P}_3^{\mu \nu}=p^{\mu} \gamma^{\nu},
\quad \mathcal{P}_4^{\mu \nu}=\gamma^{\mu} p^{\nu},\quad\mathcal{P}_5^{\mu \nu}=p^{\mu} p^{\nu},\nonumber\\
&\mathcal{P}_6^{\mu \nu}=\slashed{p} g^{\mu \nu},\quad\mathcal{P}_7^{\mu \nu}=\slashed{p} \gamma^{\mu}\gamma^{\nu},\quad \mathcal{P}_8^{\mu \nu}=\slashed{p} p^{\mu}\gamma^{\nu},
\quad \mathcal{P}_9^{\mu \nu}=\slashed{p} \gamma^{\mu} p^{\nu},\quad \mathcal{P}_{10}^{\mu \nu}=\slashed{p} p^{\mu}p^{\nu}.
\label{sp}
\end{aligned}
\end{equation}
   From Eqs.~(\ref{c})--(\ref{sp}), in the vicinity of the pole the full Delta propagator in $n$ space-time dimensions reads
\cite{Hacker:2005fh}
\begin{equation}
\begin{aligned}
iS^{\mu \nu}(p)=&-i\Bigg[g^{\mu \nu}-\frac{\gamma^ \mu \gamma^ \nu}{n-1}+\frac{p^\mu \gamma^ \nu-\gamma^ \mu p^\nu}{(n-1)z_{\Delta}}-\frac{(n-2) p^\mu p^\nu}{(n-1)z_{\Delta}^2} \Bigg]\\
&\times \frac{1}{\slashed{p}-z_{\Delta}-\Sigma_1(p^2)-\slashed{p}\Sigma_6(p^2)}+\text{pole-free terms}.
\label{hp}
\end{aligned}
\end{equation}
   The complex position of the pole of the Delta propagator, $z$, is used to define the physical pole mass and the width of the Delta resonance
\cite{Djukanovic:2007bw}, and is determined from Eq.~(\ref{hp}) by solving the equation
\begin{equation}
z-z_\Delta+4c_1^{\Delta}M^2-\tilde{\Sigma}_1(z^2)-z\Sigma_6(z^2)=0,
\label{sp2}
\end{equation}
where we have split $\Sigma_1$ into the constant tree-level piece $-4c_1^{\Delta}M^2$
and the loop contribution $\tilde{\Sigma}_1$.
\begin{figure}[t]
\includegraphics[width=\textwidth]{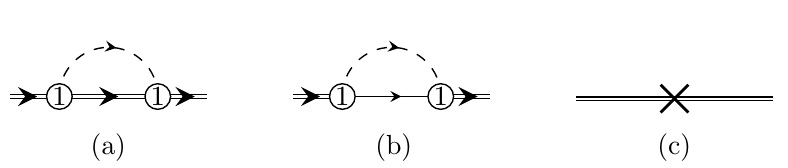}
\caption{\label{DeltaDiags1}
One-loop self-energy diagrams of the Delta resonance. The dashed, solid, and double lines correspond to the pion, nucleon and Delta, respectively.}
\end{figure}
    The contributions to the Delta self energy up to and including order $\mathcal{O}(q^3)$ from loop diagrams are shown in Fig.~\ref{DeltaDiags1}.
   Using the definition
\begin{equation}
\label{lagrangian}
\big\{I_{m_{i}}^{\mu \nu},I_{m_{i}}^{\mu \nu \lambda}\big\}
:=i\mu^{4-n} \int \frac{d^nk}{(2\pi)^n}\frac{\{k^{\mu}k^{\nu}, k^{\mu}k^{\nu}k^{\lambda}\}}{[(k-p)^2-m_i^2+i0^+](k^2-M^2+i0^+)},
\end{equation}
the unrenormalized results for loop diagrams from Fig.~\ref{DeltaDiags1} (a) and (b) are obtained as
\begin{align}
\label{dia}
\Sigma_{(a)}^{\mu \nu}=& -\frac{5\texttt{g}_{A\Delta}^2}{12F^2} \bigg\{\frac{2}{3}I_{z_\Delta}^{\mu \nu} \Big(2z_{\Delta}+3 \slashed{p}-\frac{p^2 \slashed{p}}{z_{\Delta}^2} \Big)
-g^{\mu \nu}I_{z_\Delta}^{\alpha \beta}\big[g_{\alpha \beta}(z_\Delta+\slashed{p})-2 \gamma_{\alpha}p_{\beta}\big] \\
&+g^{\mu \nu}I_{z_\Delta}^{\alpha \beta \lambda}g_{\alpha \beta}\gamma_{\lambda}
-\frac{2}{3}I_{z_\Delta}^{\mu \nu \lambda} \frac{\gamma_{\lambda}(z_{\Delta}^2-p^2)+2p_{\lambda}(z_\Delta+\slashed{p})}{z_{\Delta}^2}\bigg\},\nonumber\\
\label{dib}
\Sigma_{(b)}^{\mu \nu}=&-\frac{\texttt{g}_{AN\Delta}^2}{4F^2}\left[I_m^{\mu \nu}(\slashed{p}+m)+I_m^{\mu \nu \lambda}\gamma_\lambda\right].
\end{align}

    To carry out the complex renormalization scheme, the contributions of the self-energy loop diagrams are reduced to scalar integrals and expanded in powers of
\begin{equation*}
M^2=\mathcal{O}(q^2),\quad \slashed{p}-z_\Delta= \mathcal{O}(q),\quad p^2-z_{\Delta}^2= \mathcal{O}(q),
\end{equation*}
where $q$ denotes a small quantity.
   All terms violating the power counting are subtracted.
   With the definition
\begin{equation}
\Sigma^{(i)}(z)=\tilde{\Sigma}_1^{(i)}(z^2)+z\Sigma_6^{(i)}(z^2),\quad i=a,b,
\end{equation}
the position of the pole of the Delta resonance up to and including third order is given by
\begin{align}
\label{polD}
z=z_\Delta-4c_1^{\Delta}M^2+\big(\Sigma^{(a)}(z_\Delta)+\Sigma^{(b)}(z_\Delta)\big)-\big(\Sigma^{(a)}_\text{sub}+\Sigma^{(b)}_\text{sub}\big).
\end{align}
   In the loop contribution we have replaced $z$ by $z_\Delta$, because the difference is of ${\cal O}(M^2)$ and, thus,
results in a higher-order term.
   Expanding the integrands of the loop diagrams around $\slashed{p}=z_{\Delta}$ in powers of
$M^2$, $\slashed{p}-z_\Delta$, and $p^2-z_{\Delta}^2$ \cite{Fuchs:2003qc,Djukanovic:2009gt},
we identify all terms which are of order two or lower.
   The resulting subtraction terms read
\begin{align}
\label{suba}
\Sigma^{(a)}_\text{sub}=&\frac{5\texttt{g}_{A\Delta}^2 z_{\Delta}}{10368\pi^2 F^2}\bigg[25{z_\Delta}^2-28M^2+12\big(11z_{\Delta}^2+10M^2\big)
\text{ln}\bigg(\frac{z_\Delta}{\mu}\bigg)\bigg],\\
\label{subb}
\Sigma^{(b)}_\text{sub}=&\frac{\texttt{g}_{AN\Delta}^2}{9216\pi^2F^2z_{\Delta}^5}\bigg\{-6i\pi(z_\Delta-m)^3(z_\Delta+m)^5+C_1\nonumber\\
&-\alpha_1(m,z_\Delta)\text{ln}\bigg(\frac{z_\Delta^2-m^2}{m^2}\bigg)
 +\alpha_1(z_\Delta,m)\text{ln}\bigg(\frac{z_\Delta^2-m^2}{\mu^2}\bigg)\nonumber\\
&+M^2\bigg[12i\pi(z_\Delta-m)(z_\Delta+m)^3(2z_{\Delta}^2-z_\Delta m+2m^2)+C_2\nonumber\\
&+\alpha_2(m,z_\Delta)\text{ln}\bigg(\frac{z_\Delta^2-m^2}{m^2}\bigg)-\alpha_2(z_\Delta,m)\text{ln}\bigg(\frac{z_\Delta^2-m^2}{\mu^2}\bigg)\bigg]\bigg\},
\end{align}
with the definitions
\begin{align*}
C_1=&z_{\Delta}^2\left(-10z_{\Delta}^6-20z_{\Delta}^5m+14z_{\Delta}^4m^2+48z_{\Delta}^3m^3+9z_{\Delta}^2m^4-12z_{\Delta}^2m^5-6m^6\right),\\
C_2=&4z_{\Delta}^2\left(7z_{\Delta}^4+9z_{\Delta}^3m+3z_{\Delta}^2m^2+9z_{\Delta}m^3+6m^4\right),
\end{align*}
and
\begin{align*}
\alpha_1(m_1,m_2)=&6m_1^5\left(m_1^3+2m_1^2m_2-2m_1m_2^2-6m_2^3\right),\\
\alpha_2(m_1,m_2)=&12m_1^5\left(2m_1+3m_2\right).
\end{align*}
   Note that both subtraction terms are complex.
   The non-analytic terms {\bf $\propto M^3$} originating from Eq.~(\ref{polD}) are consistent with those of Refs.~\cite{Hacker:2005fh,Bernard:2005fy}.
   The subtraction terms of Eq.~(\ref{suba}) and Eq.~(\ref{subb}) are analytic in $M^2$ and thus in the quark mass.
   They can be absorbed in the renormalization of $z_{\Delta}$ and $c_{1}^\Delta$.
   Both constants are complex and remain finite in the limit $m \rightarrow z_{\Delta}$.

   To show that the renormalized diagrams have the chiral order $\mathcal{O}(q^3)$ and, thus, satisfy the power-counting scheme,
we first consider the diagram (a).
   Dividing the contribution by $M^3$, we obtain in the limit $M\rightarrow 0$
\begin{equation}
\lim_{M\to 0} \frac{1}{M^3}\Sigma^{(a)}_{\text{ren}}(z_\Delta)=\frac{25\texttt{g}_{A\Delta}^2}{864\pi F^2}.
\end{equation}
   For a constant and finite mass difference $z_{\Delta}-m$, the contribution of diagram (b) divided by $M^3$ gives zero in the limit $M\rightarrow 0$:
\begin{equation}
\lim_{M\to 0} \frac{1}{M^3}\Sigma^{(b)}_{\text{ren}}(z_\Delta)=0.
\end{equation}
   If we scale the mass difference $z_{\Delta}-m$ as $\alpha M$ \cite{Djukanovic:2009gt}, we find the following result as $M\to 0$,
\begin{equation}
\lim_{M\to 0} \frac{1}{M^3}\Sigma^{(b)}_{\text{ren}}(z_\Delta)=\frac{\texttt{g}_{AN\Delta}^2}{192\pi^2F^2}f(\alpha),
\end{equation}
with
\begin{align}
f(\alpha)&=\alpha+(6\alpha-4\alpha^3)\text{ln}(2\alpha)+4(\alpha^2-1)^\frac{3}{2}\text{ln}\big(\sqrt{\alpha^2-1}+\alpha\big)\nonumber\\
&\quad+4i\pi\alpha^3-6i\pi\alpha+4i\pi(1-\alpha^2)\sqrt{\alpha^2-1}.
\end{align}
   Finally, first taking the limit $z_{\Delta}\rightarrow m$ and then $M\rightarrow 0$, we obtain
\begin{equation}
\lim_{M\to 0} \frac{1}{M^3}\lim_{z_\Delta\to m}\Sigma_{(b)}^{\text{ren}}(z_\Delta)=-\frac{\texttt{g}_{AN\Delta}^2}{96F^2\pi}.
\end{equation}
   In other words, even if the mass difference $z_\Delta-m$ is counted as ${\cal O}(q)$ like in the
small-scale expansion (SSE) \cite{Hemmert:1997ye}, the contributions of the renormalized self-energy diagrams behave in the vicinity of the
Delta pole as  $\mathcal{O}(q^3)$.

   Writing the position of the pole as
\begin{displaymath}
z=\text{Re}(z)+i\text{Im}(z)=m_{\Delta\,\text{pole}}-\frac{i}{2}\Gamma,
\end{displaymath}
we obtain the expressions
\begin{align}
\label{mDeltaexpansion}
m_{\Delta\,\text{pole}}&=\text{Re}(z_\Delta)-4\text{Re}(c_1^\Delta)M^2+m_{\Delta\,\text{pole}}^{\text{loop}},\\
\label{Gammaexpansion}
\Gamma&=-2\text{Im}(z_\Delta)+8\text{Im}(c_1^\Delta)M^2+\Gamma^{\text{loop}},
\end{align}
where $z_\Delta$ refers to the chiral limit of the pole.
   Note that the expressions in the chiral limit are free parameters of the theory.
   Moreover, since also $c_1^\Delta$ is unknown, we first provide an order-of-magnitude estimate
by assuming $\text{Re}(c_1^\Delta)\approx -1$ GeV$^{-1}$, resulting in a contribution to
$m_{\Delta\,\text{pole}}$ of approximately 78 MeV.
   To discuss the contributions of the loop diagrams we make use of the numerical values $m_N=m_p=938.3$ MeV,
$M_\pi=M_{\pi^+}=139.6$ MeV, $z_\Delta=(1210-i50)$ MeV, $F_\pi=92.2$ \text{MeV}, $g_A=1.27$, $g_{A\Delta}=\frac{9}{5}g_A=0.706$,
and $g_{AN\Delta}=\frac{6}{5}\sqrt{2}g_A=2.16$.
   The last two values are obtained from the static SU(6) quark model.
   The loop results are then given by
\begin{align*}
m_{\Delta\,\text{pole}}^{\text{loop}}&=13.8\,\text{MeV}=(12.2+1.6)\,\text{MeV},\\
\Gamma^{\text{loop}}&=2.26\,\text{MeV}=(0.10+2.16)\,\text{MeV},
\end{align*}
where we split the total results into the contributions of diagram (a) and (b), respectively.

   In the following, we compare the numerical values of the renormalized loop diagrams with the SSE \cite{Hemmert:1997ye},
where the mass difference $\delta\equiv z_\Delta-m$ and the lowest-order pion mass $M$ both count as ${\cal O}(q)$.
   To that end, we substitute $M\to t M$ and $\delta \to t \delta$ in the loop expressions, expand the result in $t$, subtract the
terms proportional to $t^0$, $t$, and $t^2$, and, in the end, set $t=1$ in the resulting expressions.
   Whatever is left over, we call the renormalized SSE loop contribution at ${\cal O}(q^3)$.
   The strategy is the same as in the extended on-mass-shell scheme \cite{Fuchs:2003qc}, and the renormalized
result also keeps higher-order terms of the loop contribution.
   The difference between the heavy-baryon approach \cite{Jenkins:1990jv,Bernard:1992qa,Jenkins:1991es,Banerjee:1994bk,Hemmert:1997ye}
and the covariant approach \cite{Tang:1996ca,Ellis:1997kc,Becher:1999he,Gegelia:1999gf,Gegelia:1999qt,Fuchs:2003qc,Pascalutsa:2005nd,Bernard:2003xf,Hacker:2005fh,Bernard:2005fy}
is that the latter includes an infinite series of relativistic corrections of higher orders.
   Moreover, with the above method we may also identify the terms of a given loop contribution that scale as $t^n$, $n\geq 3$.
   We denote the $t^3$ terms as SSE-LO and the sum of the $t^3$ and $t^4$ terms as SSE-NLO, respectively.
   The comparison with the full result, to some extent, allows us to estimate the uncertainty due to higher-order
terms.
   In the SSE, the power-counting-violating terms are obtained as
\begin{align}
\Sigma^{(a)}_\text{sub}=&\frac{5\texttt{g}_{A\Delta}^2 z_{\Delta}}{10368\pi^2 F^2}
\bigg[25z_\Delta^2-28M^2+12\big(11z_{\Delta}^2+10M^2\big)\text{ln}\bigg(\frac{z_\Delta}{\mu}\bigg)\bigg]\nonumber,\\
\label{sse}
\Sigma^{(b)}_\text{sub}=&-\frac{\texttt{g}_{AN\Delta}^2z_\Delta}{9216\pi^2F^2}\bigg\{-23z_\Delta^2+32z_\Delta\delta+8(26\delta^2-17M^2)\nonumber\\
&+60\left[z_\Delta^2-4z_\Delta\delta+2(M^2+2\delta^2)\right]\text{ln}\bigg(\frac{z_\Delta}{\mu}\bigg)\bigg\},
\end{align}
where, in the SSE, the renormalized mass parameter $z_\Delta$ is chosen to be real.
   The SSE-LO result for the imaginary part of the Delta pole position $z$ is then given by \cite{Hacker:2005fh}
\begin{equation}
\text{Im}(z)=-\frac{\texttt{g}_{AN\Delta}^2(\delta^2-M^2)^{\frac{3}{2}}}{48\pi F^2}+\mathcal{O}(q^4).
\label{imzsse}
\end{equation}
   As $M$ and $\delta$ are of ${\cal O}(q)$, $\text{Im}(z)$ of Eq.~(\ref{imzsse}) is of
$\mathcal{O}(q^3)$ and, thus, satisfies the corresponding power counting.

\begin{figure}[t]
\includegraphics[width=\textwidth]{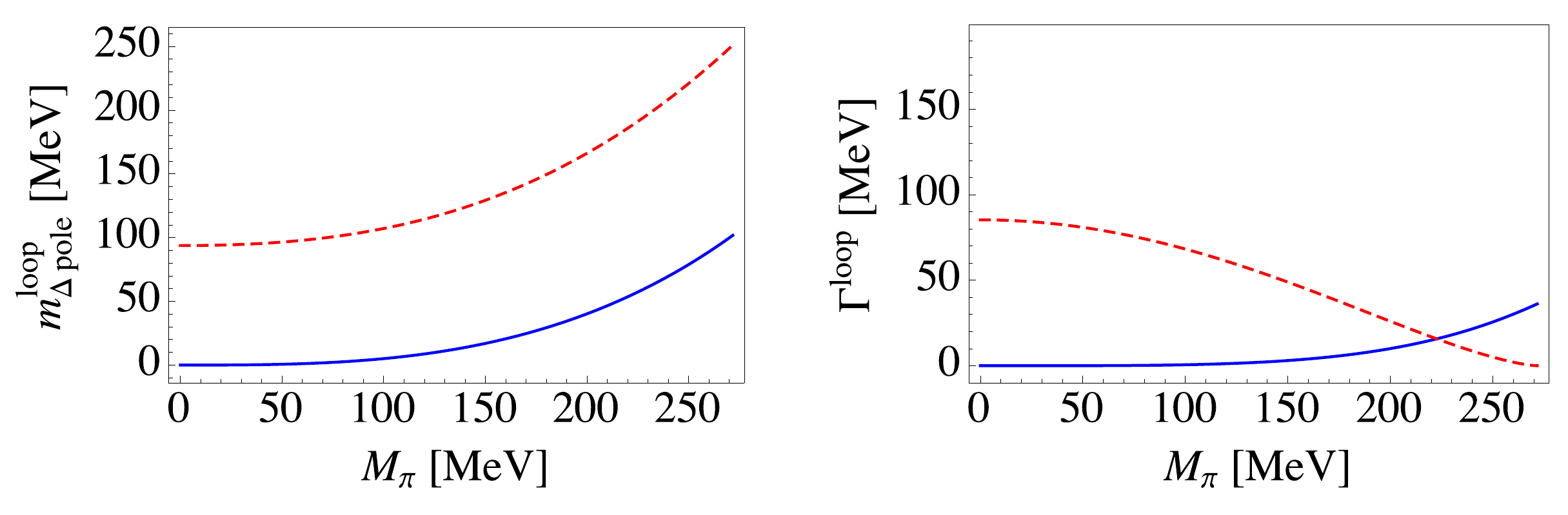}
\caption{(Color online) Contributions of the renormalized loop diagrams to the real part of the Delta-pole position
(left panel) and the width (right panel) as a function of the pion mass.
The solid (blue) lines denote the CMS and the dashed (red) lines the SSE.
\label{comparison_CMS_SSE}}
\end{figure}

   In Fig.~\ref{comparison_CMS_SSE}, we display the renormalized loop contributions $m_{\Delta\,\text{pole}}^{\text{loop}}$ and
$\Gamma^{\text{loop}}$ as functions of the pion mass $M=M_\pi$ for the CMS [solid (blue) curves] and the SSE [dashed (red) curves].
   In the CMS we have implemented the renormalization condition that $z_\Delta$ be the position of the pole
in the chiral limit.
   For that reason, the renormalized loop contribution vanishes in the chiral limit in this scheme, i.e., both
$m_{\Delta\,\text{pole}}^{\text{loop}}$ and $\Gamma^{\text{loop}}$ vanish $\propto M^3$ as $M\to 0$.
   The renormalized loop contributions to the real part of the pole are compatible with Refs.~\cite{Bernard:2005fy,Bernard:2009mw},
where the Delta mass of fourth order has been analyzed in the SSE using lattice data and chiral effective field theory.

\begin{figure}[t]
\includegraphics[width=\textwidth]{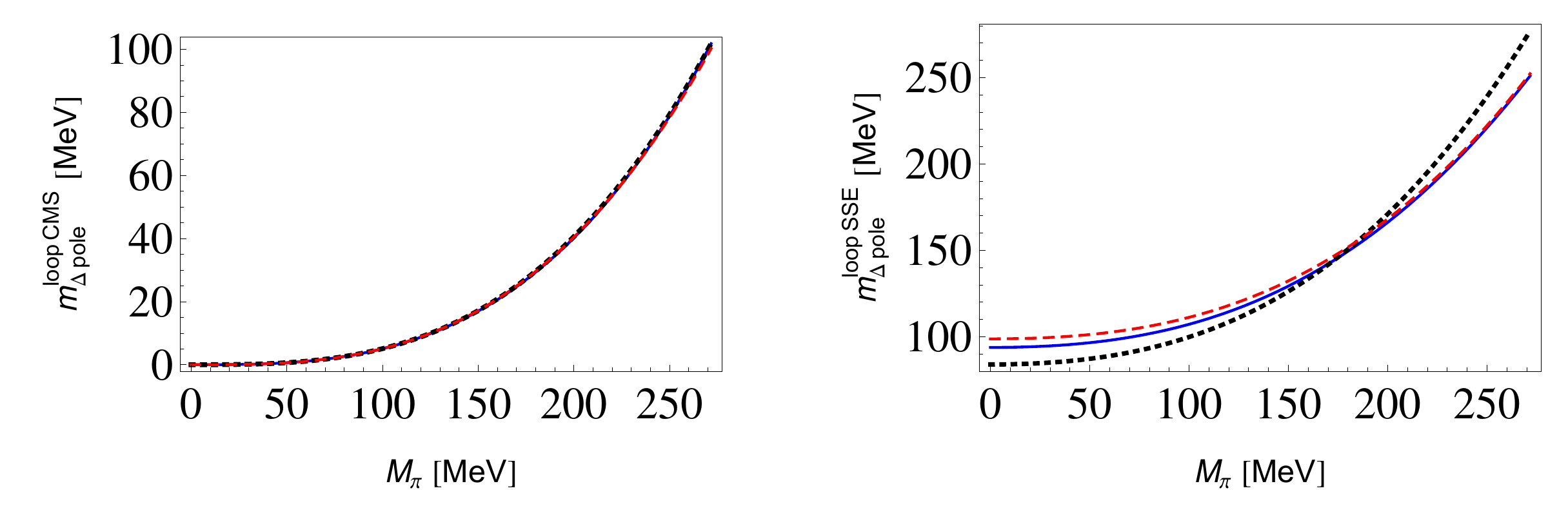}
\caption{(Color online)
Pion-mass dependence of the loop contribution to the real part of the Delta pole.
Left panel: CMS. Right panel: SSE. Solid (blue) curves denote the full result, dotted (black) curves
the LO expansion, and dashed (red) curves the NLO expansion, respectively. In the CMS case the curves are essentially
indistinguishable.
\label{comparison_expansion_mass_cms_sse}}
\end{figure}

\begin{figure}[t]
\includegraphics[width=\textwidth]{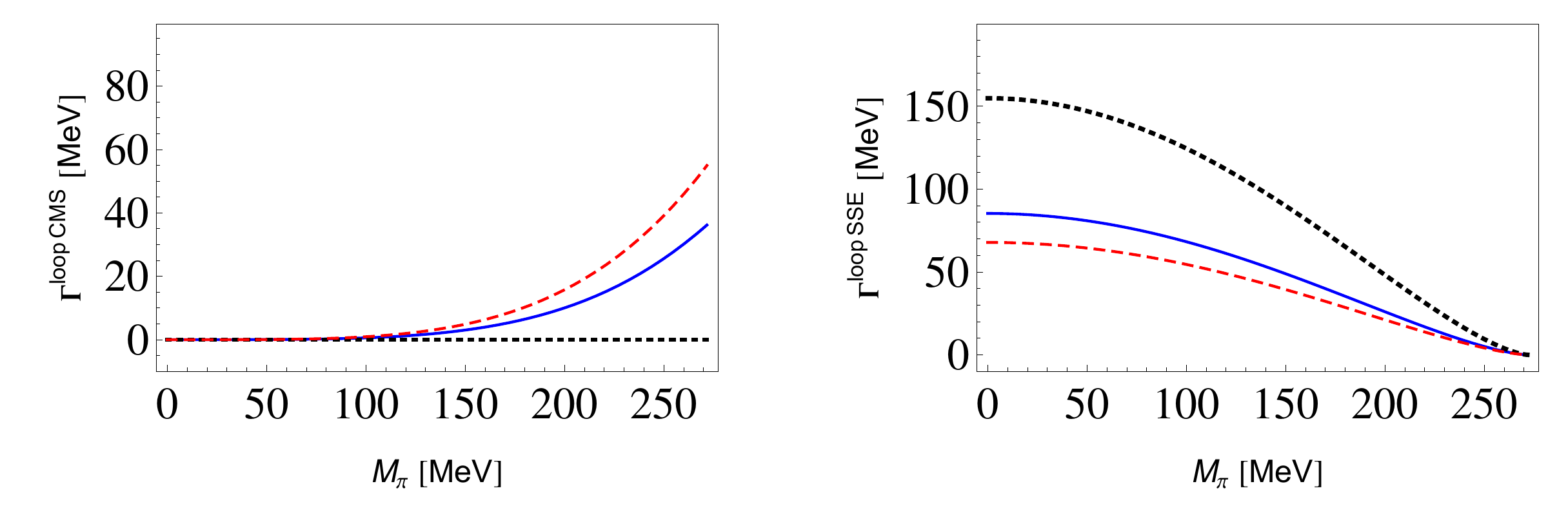}
\caption{(Color online)
Pion-mass dependence of the loop contribution to the width.
Left panel: CMS. Right panel: SSE. Solid (blue) curves denote the full result, dotted (black) curves
the LO expansion, and dashed (red) curves the NLO expansion, respectively.
\label{comparison_expansion_width_cms_sse}}
\end{figure}

   In Figs.~\ref{comparison_expansion_mass_cms_sse} and \ref{comparison_expansion_width_cms_sse}, we display the pion-mass
dependence of the renormalized loop contributions $m_{\Delta\,\text{pole}}^{\text{loop}}$ and
$\Gamma^{\text{loop}}$, respectively.
   To identify the higher-order contributions contained in the loop results, in each case we explicitly
show the LO result [dotted (black) curve], the NLO result [dashed (red) curve], and the full result [solid (blue) curve].
   Our first observation is that, in the CMS, the loop contribution to the real part of the Delta pole is essentially
already given by the leading-order term, in other words, the three curves are hardly distinguishable.
   On the other hand, for the same quantity, small differences are visible between the different orders
in the SSE.
   Regarding the width, the situation is different.
   In the CMS, the leading-order contribution to $\Gamma^{\text{loop}}$ vanishes, and the three curves start
to deviate from each other roughly at 100 MeV.
   Furthermore, in the SSE, the LO result substantially differs from the full result, while the
NLO result is, as expected, already closer to the full result.
   When discussing the width, two features need to be kept in mind.
   First, in the SSE, the width is entirely given by the loop contribution, whereas in the CMS we expect a substantial
contribution from $\Gamma_\chi$ and a smaller contribution from the $c_1^\Delta$ term.
   Second, the power counting of the SSE assumes that the lowest-order pion mass $M$ and the mass difference
$\delta=m_\Delta-m$ scale in the same way.
   In that sense, taking the limit $M\to 0$ and keeping the mass difference $\delta=m_\Delta-m$ fixed,
is in some way against the spirit of the SSE.
   In general, we observe that neglecting higher-order contributions has a greater impact on the calculation
in the SSE when considering pion masses $\leq$ $M_{\pi^+}$.

\begin{figure}[t]
\includegraphics[width=0.5\textwidth]{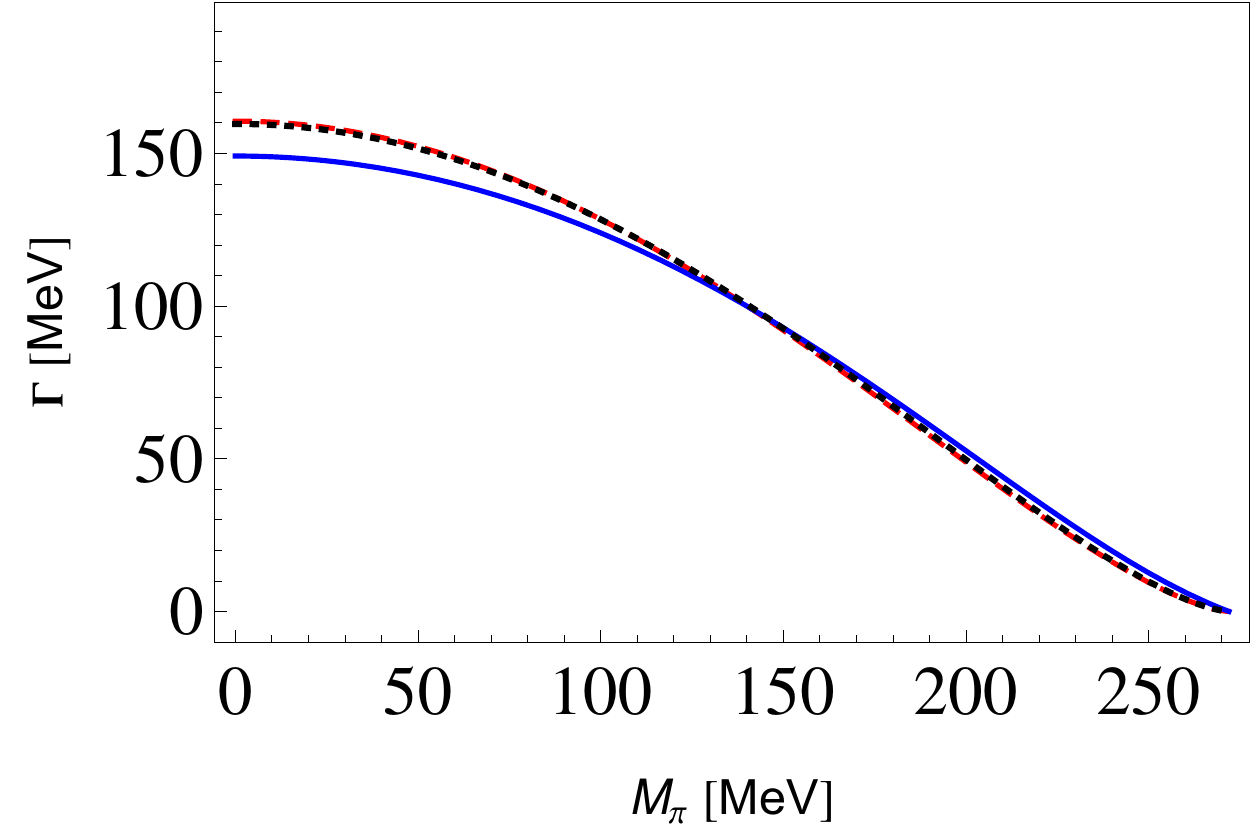}
\caption{(Color online) Pion-mass dependence of the width.  The solid (blue) curve denotes the CMS result
of Eq.~(\ref{GammaM}), the dashed (red) curve denotes the SSE result with $g_{AN\Delta}=3.04$, and the dotted (black)
curve denotes the LO-SSE result with $g_{AN\Delta}=2.25$.
   In each case the parameters are chosen such that the width at the physical mass $M_{\pi^+}=139.6$~MeV
reproduces the experimental value of $\Gamma_{\text{exp}}=100$~MeV.
\label{comparisons_full_widths}}
\end{figure}

    Next, we extract estimates for the width in the chiral limit,
$\Gamma_\chi$, and for the parameter $\text{Im}(c_1^\Delta)$.
   To that end, we interpret Eq.~(\ref{Gammaexpansion}) as a function of the
pion mass $M$,
\begin{equation}
\label{GammaM}
\Gamma(M)=\Gamma_\chi+8\text{Im}(c_1^\Delta) M^2+\Gamma^{\text{loop}}(M).
\end{equation}
   At $M=M_\pi=M_{\pi^+}$, we adjust the width to the empirical number $\Gamma(M_\pi)=\Gamma_{\text{exp}}=100$ MeV.
When $M$ reaches the mass difference $\delta=m_\Delta-m_N$, the Delta resonance is stable
and the width is $\Gamma(\delta)=0$.
   We thus have two equations in two unknowns,
\begin{align*}
\Gamma_\chi+8M_\pi^2\,\text{Im}(c_1^\Delta)&=\Gamma_{\text{exp}}-\Gamma^{\text{loop}}(M_\pi),\\
\Gamma_\chi+8\delta^2\,\text{Im}(c_1^\Delta)&=-\Gamma^{\text{loop}}(\delta),
\end{align*}
with the solution
\begin{align*}
\begin{pmatrix}
\Gamma_\chi\\
\text{Im}(c_1^\Delta)
\end{pmatrix}
&=\frac{1}{8(\delta^2-M_\pi^2)}
\begin{pmatrix}
8\delta^2\left[\Gamma_{\text{exp}}-\Gamma^{\text{loop}}(M_\pi)\right]+8M_\pi^2\Gamma^{\text{loop}}(\delta)\\
-\Gamma_{\text{exp}}+\Gamma^{\text{loop}}(M_\pi)-\Gamma^{\text{loop}}(\delta)
\end{pmatrix}.
\end{align*}
   For $g_{AN\Delta}=2.16$, we obtain $\Gamma_\chi=149$ MeV and $\text{Im}({c_1^\Delta})=-0.313$~GeV$^{-1}$.
   In Fig.~\ref{comparisons_full_widths}, we compare
three scenarios, namely, the full CMS result [solid (blue) curve],
the full SSE result [dashed (red) curve], and the LO-SSE result [dotted (black) curve].
   All results are normalized such that they produce the value $\Gamma_{\rm exp}$ at the physical
pion mass $M_{\pi^+}=139.6$~MeV.
   For the full SSE result this requires the value $g_{AN\Delta}=3.04$.
   In this case, we obtain for the width in the chiral limit $\Gamma_\chi=161$~MeV.
   On the other hand, in terms of the LO-SSE result, the width is reproduced for $g_{AN\Delta}=2.25$
   with a width in the chiral limit of $\Gamma_\chi=160$~MeV.
   Note that, by construction, the results must agree at $M=0$ and $M=M_{\pi^+}$.
   From Fig.~\ref{comparisons_full_widths} we draw two main conclusions.
   First, the CMS-width function differs only slightly from the SSE-width function, where
the difference is largest for small values of the pion mass.
   Furthermore, below (above) the physical pion mass, the CMS width is smaller (larger) than the SSE width.
   Second, the difference between the full SSE result and the LO SSE from Fig.~\ref{comparison_expansion_width_cms_sse}
(right panel) can be (almost) completely compensated by adjusting the LEC $g_{AN\Delta}$.
   As a consequence, the dashed (red) curve and the dotted (black) curve almost coincide in Fig.~\ref{comparisons_full_widths}.

\begin{figure}[t]
\includegraphics[width=\textwidth]{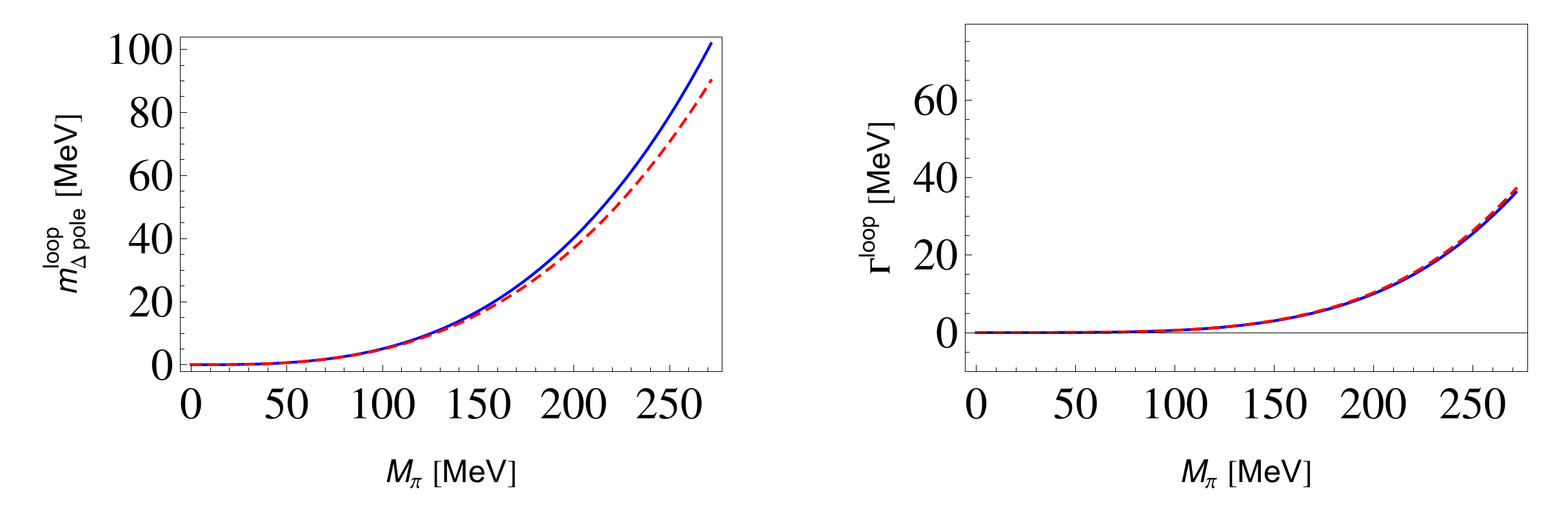}
\caption{(Color online) Contributions of the renormalized loop diagrams to the real part of the Delta-pole position
(left panel) and the width (right panel) as a function of the pion mass.
The solid (blue) lines denote the CMS result with the Lagrangian of Ref.~\cite{Wies:2006rv} [see
Eqs.~(\ref{LpiDelta1}) and (\ref{LpiNDelta1})],
and the red (dashed) lines denote the CMS result with the Lagrangian of Ref.~\cite{Gegelia:2016pjm}
[see Eq.~(\ref{alternative_Lagrangians})].
\label{comparison_two_lagrangians}}
\end{figure}

   Finally, in Fig.~\ref{comparison_two_lagrangians}, we compare our results for the renormalized loop contributions
$m_{\Delta\,\text{pole}}^{\text{loop}}$ and $\Gamma^{\text{loop}}$ [solid (blue) curves]
with those obtained in the framework of Ref.~\cite{Gegelia:2016pjm} [dashed (red) curves].
   Our approach is based on the Dirac constraint analysis of Ref.~\cite{Wies:2006rv} and leads to
manifestly consistent interactions with the correct number of dynamical degrees of freedom.
   On the other hand, Ref.~\cite{Gegelia:2016pjm} fixes the off-shell parameters such that the
interaction Lagrangians are given by the sum
\begin{equation}
\label{alternative_Lagrangians}
{\cal L}^{(1)}_{\pi\Delta\Delta}+{\cal L}^{(1)}_{\pi N\Delta}=
\frac{1}{2}\frac{\texttt{g}_{A\Delta}}{F}\bar{\Psi}_{\mu} \xi^{\frac{3}{2}}\gamma^\rho\gamma_5\vec\tau\cdot\partial_\rho\vec\pi g^{\mu\nu}
\xi^{\frac{3}{2}}\Psi_\nu
+\left(-\frac{1}{2}\frac{\texttt{g}_{AN\Delta}}{F}\bar{\Psi}_{\mu, i}\;\xi_{ij}^\frac{3}{2}g^{\mu \nu}
\partial_\nu\pi_j\Psi+\text{h.c.}\right).
\end{equation}
   The reasoning for this approach is that the effects of off-shell parameters can be absorbed in
LECs of other terms of the effective Lagrangian \cite{Gegelia:2016pjm}.
   For the loop contribution to the real part of the Delta-pole position we see a small difference
between the two calculations, which increases with the pion mass.
   Clearly, this difference may be regarded as a higher-order effect.
   Furthermore, the results for the widths are almost indistinguishable.

\section{Summary and conclusions}

   We have discussed the pole position of the $\Delta(1232)$ resonance within the framework of covariant
chiral effective field theory.
   For the interaction terms, we made use of the results of a Dirac constraint analysis.
   To implement a consistent power-counting scheme, we investigated both the complex-mass renormalization scheme
and the covariant small-scale expansion combined with the extended on-mass-shell scheme.
   For both renormalization schemes, we explicitly identified the power-counting-violating terms.
   In comparison with the heavy-baryon expansion, the covariant calculations at ${\cal O}(q^3)$ involve the
summation of an infinite series of relativistic corrections.
   We investigated the renormalized loop contributions to the complex pole position with respect to
their pion-mass dependence (see Fig.~\ref{comparison_CMS_SSE}).
   As a consequence of the renormalization condition, in the CMS both $m_{\Delta\,\text{pole}}^{\text{loop}}$ and
$\Gamma^\text{loop}$ vanish as $M\to 0$.
   On the other hand, in the SSE both quantities approach non-vanishing constants in this limit.
   To judge the importance of the higher-order terms implicit in the loop integrals,  we also provided,
in addition to the full loop contributions, the LO and NLO results with respect to a rescaling of the
small quantities (see Figs.~\ref{comparison_expansion_mass_cms_sse} and \ref{comparison_expansion_width_cms_sse}).
   Moreover, we discussed the function for the width within three different scenarios (see Fig.~\ref{comparisons_full_widths}).
   In the CMS, we took $g_{AN\Delta}=2.16$ and determined $\Gamma_\chi=149$ MeV and $\text{Im}(c_1^\Delta)=-0.313$ GeV$^{-1}$
from $\Gamma(M_{\pi^+})=\Gamma_{\text{exp}}=100$~MeV and $\Gamma(\delta)=0$.
   In the SSE, the width is completely given by the loop contribution of Fig.~1~(b).
   We fixed the parameter $g_{AN\Delta}$ to the experimental value $\Gamma_\text{exp}$ at $M_{\pi^+}$, resulting
in $g_{AN\Delta}=3.04$ when using the full SSE expression, and in $g_{AN\Delta}=2.25$ for the LO-SSE result.
   The corresponding values for the width in the chiral limit read 161 MeV (full SSE expression) and
160 MeV (LO-SSE expression), respectively.
   In the SSE, the width automatically vanishes at $M=\delta$.
   Unfortunately, we cannot completely fix the pion-mass dependence of the real part
of the pole position.
   The reason is that in both cases we still have two free parameters, namely, $m_\Delta$ and $\text{Re}(c_1^\Delta)$ in the CMS,
and $\delta=m_\Delta-m$ and $c_1^\Delta$ in the SSE.
   On the other hand, there is only one condition, namely, the real part of the pole position, 1210 MeV, at the physical
pion mass $M_{\pi^+}$.
   Finally, we compared the CMS results of two interaction Lagrangians differing in their
``off-shell'' behavior [see Fig.~\ref{comparison_two_lagrangians}].
   The differences in the loop contributions turn out to be small and
may be regarded as higher-order effects beyond the accuracy of our calculation.

\section{Acknowledgments}
   The authors would like to thank J.~Gegelia and D.~Djukanovic for valuable contributions to the work.
   The work of T.~B.~was supported by a scholarship of the German Academic Exchange Service.
   T.~B.~would like thank H.~W.~Grie{\ss}hammer for the warm hospitality he received during his stay at
The George Washington University, where part of this work was done.
   The work of Y.~\"U.~was supported by the Scientific and Technological Research Council of Turkey (T\"UB\.{I}TAK).

\begin{appendix}

\section{Feynman rules}
   The Feynman rules for the propagators and vertices are given in Table \ref{table_feynman_rules}.
\begin{center}
\begin{table}[ht]
\caption{Feynman rules for propagators and vertices. Note that
$i, j, l$ and $r, s$ correspond to cartesian isospin triplet and isospin doublet
indices, respectively.\label{table_feynman_rules}}
\centering
\begin{tabular}{l c c c}
\hline\hline
& Propagators & Vertices \\ [0ex]
\hline \\[0.5ex]
&
  \pbox{\textwidth}{
  \begin{tikzpicture}[scale=1]
  \draw [dashed,,->-=.55] (-1,0) -- (1,0);
  \node [below=3pt]{$k$};
  \draw [fill] (-1,0) circle (.05)node [anchor=east]{$i$};
  \draw [fill] (1,0) circle (.05)node [anchor=west]{$j$};
  \end{tikzpicture}
}
&
  \pbox{\textwidth}{
  \begin{tikzpicture}[scale=1]
  \draw [double,->-=.55] (-1,0) -- (0,0);
  \node [below left] at(-.5,-.05) {$\psi_i^{\alpha}(p)$};
  \draw [double,-<-=.55] (1,0) -- (0,0);
  \node [below right] at(.5,.01) {$\psi_j^{\beta}(p')$};
  \draw [dashed,-<-=.55] (0,0) -- (0,1);
  \node [above right] at(.08,.3) {$\phi^l(k)$};
  \draw [fill=white] (0,0) circle(.15) node [anchor=center] {$\scriptstyle 1$};
    \end{tikzpicture}
} \\  [8ex]
&
$  \frac{i}{k^2-M^2+i\epsilon}\delta_{ij}$ & $\begin{array} {lcl}
 & \frac{\texttt{g}_{A\Delta}}{2F}
   \left[g^{\alpha \beta} \gamma^\mu k_\mu \gamma_5-(\gamma^\beta k^\alpha+\gamma^\alpha k^\beta)\gamma_5-\gamma^\beta \gamma^\mu \gamma_5 \gamma^\alpha k_\mu\right]\\
 & \times\left(\frac{5}{9}i\epsilon_{jil}+\frac{1}{9}\delta_{il}\tau_j+\frac{1}{9}\delta_{jl} \tau_i-\frac{4}{9}\delta_{ji} \tau_l\right)\end{array}$ \\ [8ex]
&
  \pbox{\textwidth}{%
  \begin{tikzpicture}[scale=1]
  \draw [->-=.55] (-1,0) -- (1,0);
  \node [below=3pt]{$p$};
  \draw [fill] (-1,0) circle (.05)node [anchor=east]{$r$};
  \draw [fill] (1,0) circle (.05)node [anchor=west]{$s$};
  \end{tikzpicture}
}
&
  \pbox{\textwidth}{
  \begin{tikzpicture}[scale=1]
  \draw [double,->-=.55] (-1,0) -- (0,0);
  \node [below left] at(-.5,-.05) {$\psi_i^{\alpha}(p)$};
  \draw [double,-<-=.55] (1,0) -- (0,0);
  \node [below right] at(.5,.01) {$\Psi(p')$};
  \draw [dashed,-<-=.55] (0,0) -- (0,1);
  \node [above right] at(.08,.3) {$\phi^l(k)$};
  \draw [fill=white] (0,0) circle(.15) node [anchor=center] {$\scriptstyle 1$};
  \end{tikzpicture}
}
\\  [8ex]
&  $\frac{i}{\slashed{p}-m+i\epsilon}\delta_{rs}$ & $ \frac{\texttt{g}_{AN\Delta}}{2F}k_\nu(g^{\nu \alpha}-\gamma^\nu \gamma^\alpha)
   \left(\delta_{li}-\frac{1}{3}\tau_l \tau_i\right)$  \\ [8ex]
&
 \pbox{\textwidth}{
 \begin{tikzpicture}[scale=1]
 \draw [double,->-=.55] (-1,0) -- (1,0);
 \node [below=3pt]{$p$};
 \draw [fill] (-1,0) circle (.05)node [anchor=east]{$\mu,\alpha$};
 \draw [fill] (1,0) circle (.05)node [anchor=west]{$\nu,\beta$};
 \end{tikzpicture}%
}\\  [8ex]
&  $ -i\frac{\slashed{p}+m_\Delta}{p^2-m_\Delta^{2}}\big(g^{\mu \nu}-\frac{\gamma^\mu \gamma^\nu}{n-1}-\frac{\gamma^\mu p^\nu- p^\nu \gamma^\mu }{(n-1)m_\Delta}
-\frac{(n-2)p^\alpha p^\beta}{(n-1)m_\Delta^{2}} \big)$ \\ [3ex]
\\ [.5ex]
\hline\hline
\end{tabular}
\label{table:Fd}
\end{table}
\end{center}

\section{Loop integrals}
The scalar loop integrals of one-, and two-point functions which are used for the calculation of the self-energy diagrams are given by
\begin{align*}
&A_0(m^2)= \frac{(2\pi\mu)^{4-D}}{i\pi^2} \int d^Dk \frac{1}{k^2-m^2}, \\
&B_0(p_1^2, m_1^2, m_2^2)= \frac{(2\pi\mu)^{4-D}}{i\pi^2} \int d^Dk \frac{1}{[k^2-m_1^2][(k+p_1)^2-m_2^2]}.
\end{align*}
\end{appendix}

\end{document}